\documentclass{osa-article}
\journal{oe}
\articletype{Research Article}

\usepackage{upgreek}
\usepackage{gensymb}
\newcommand{\mrm}[1]{\mathrm{#1}}
\newcommand{\micron}[0]{$\upmu$m}

\begin{document}

\title{Efficient second harmonic generation in nanophotonic GaAs-on-insulator waveguides}

\author{Eric~J.~Stanton,\authormark{1,*} Jeff~Chiles,\authormark{1} Nima~Nader,\authormark{1} Galan~Moody,\authormark{1,$\dagger$} Nicolas~Volet,\authormark{2} Lin~Chang,\authormark{3} John~E.~Bowers,\authormark{3} Sae~Woo~Nam,\authormark{1} and Richard~P.~Mirin \authormark{1}}

\address{\authormark{1}Applied Physics Division, National Institute of Standards and Technology, Boulder, CO 80305, USA\\
\authormark{2}Department of Engineering, Aarhus University, 8200 Aarhus N, Denmark\\
\authormark{3}Department of Electrical and Computer Engineering, University of California, Santa Barbara, CA 93106, USA\\
\authormark{$\dagger$}Current affiliation: Department of Electrical and Computer Engineering, University of California, Santa Barbara, CA 93106, USA}

\email{\authormark{*}eric.stanton@nist.gov}

\begin{abstract*}
Nonlinear frequency conversion plays a crucial role in advancing the functionality of next-generation optical systems. Portable metrology references and quantum networks will demand highly efficient second-order nonlinear devices, and the intense nonlinear interactions of nanophotonic waveguides can be leveraged to meet these requirements. Here we demonstrate second harmonic generation (SHG) in GaAs-on-insulator waveguides with unprecedented efficiency of 40~W$^{-1}$ for a single-pass device. This result is achieved by minimizing the propagation loss and optimizing phase-matching. We investigate surface-state absorption and design the waveguide geometry for modal phase-matching with tolerance to fabrication variation. A 2.0~\micron{} pump is converted to a 1.0~\micron{} signal in a length of 2.9~mm with a wide signal bandwidth of 148~GHz. Tunable and efficient operation is demonstrated over a temperature range of 45~\celsius{} with a slope of 0.24~nm/\celsius{}. Wafer-bonding between GaAs and SiO$_2$ is optimized to minimize waveguide loss, and the devices are fabricated on 76~mm wafers with high uniformity. We expect this device to enable fully integrated self-referenced frequency combs and high-rate entangled photon pair generation.
\end{abstract*}

\section{Introduction}

Optical frequency conversion uses nonlinear interactions of light with matter to generate new wavelengths of light. Recent efforts have improved conversion efficiencies and broadened the application of this technology by using nanophotonic waveguides. Second harmonic generation (SHG) is one nonlinear process of significant interest because it is critical for stabilizing optical frequency combs \cite{Diddams2000}. In SHG, input pump light is converted to a signal with twice the frequency. It is used to self-reference a frequency comb by comparing and controlling the difference between two sections of the comb spectra that are separated by an octave. This is referred to as $f$-$2f$ self-referencing and it produces low-noise optical frequency combs used for optical clocks \cite{Udem2002,Papp2014,Drake2019}, optical frequency synthesizers \cite{Holzwarth2000,Briles2018,Spencer2018}, low-phase-noise microwave generation \cite{Fortier2011}, and molecular sensing \cite{Coddington2008,Coddington2016}.

Increasing the application space for stabilized frequency combs requires reducing the cost, size, weight, and power consumption while maintaining performance metrics \cite{Spencer2018}. Chip-scale integration is an attractive way to achieve these goals, particularly using heterogeneous integration to combine a suite of different materials into a compact package. A promising approach to develop a chip-scale stabilized frequency comb is to pump a SiN microresonator with a continuous-wave (CW) near-IR laser \cite{Li2017,Briles2018,Spencer2018}. This design is scalable to large-volume and high-yield manufacturing to reduce cost. However, exceptionally high nonlinear conversion efficiencies are required for this fully integrated system. In particular, avoiding the use of amplifiers for the microresonator pump laser and the SHG pump light will be a significant advancement \cite{Volet2018}. Given the limitations presented by the SiN microresonator quality factor that restricts the pump power and by the detector noise, extremely high SHG efficiency is necessary to achieve a sufficient signal-to-noise ratio.

\begin{figure}[htbp]
\centering\includegraphics{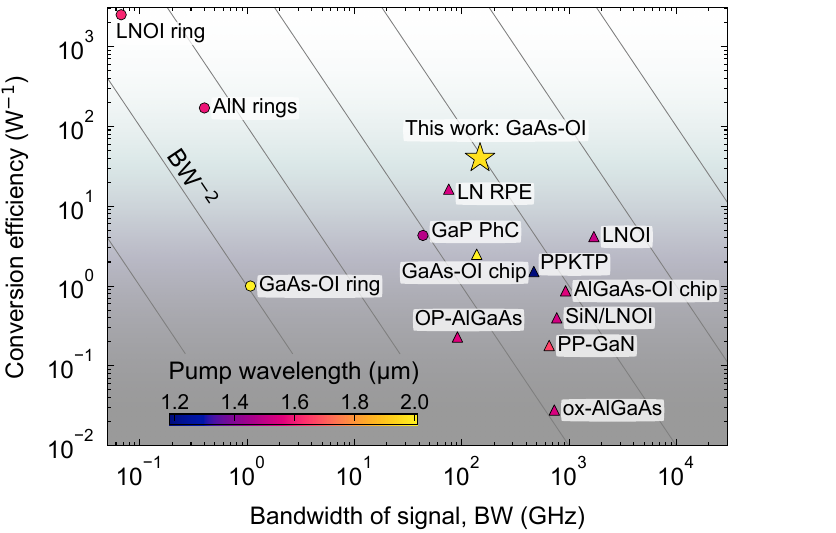}
\caption{Highest SHG conversion efficiencies demonstrated from different waveguide material platforms for resonant (circles) and single-pass (triangles) devices. The signal bandwidth is defined by the FWHM. References: AlGaAs-OI chip \cite{May2019}, AlN rings \cite{Bruch2018}, GaAs-OI chip \cite{Chang2018}, GaAs-OI ring \cite{Chang2019}, GaP PhC \cite{Rivoire2009}, LNOI \cite{Wang2018}, LNOI ring \cite{Lu2019}, LN RPE \cite{Parameswaran2002}, OP-AlGaAs \cite{Yu2005}, ox-AlGaAs \cite{Savanier2011}, PP-GaN \cite{Chowdhury2003}, PPKTP \cite{Fedorova2012}, SiN/LNOI \cite{Chang2016}.}
\label{fig:metric}
\end{figure}

A comparison of the highest SHG conversion efficiencies with the corresponding signal bandwidth from different waveguide platforms is shown in Fig.\,\ref{fig:metric} for the non-depleted pump regime. The pump wavelength is also indicated since the conversion efficiency is generally higher for shorter wavelengths with a given platform \cite{Yariv1973,Pliska1998}. Neglecting the effects of non-uniformity and loss, conversion efficiency is expected to increase inversely proportional to the square of the bandwidth. Larger SHG bandwidths are especially important for applications where the pump is either broad or consists of multiple narrow-linewidth modes that reside within the SHG bandwidth \cite{Jankowski2020}. In certain cases, only one narrow-linewidth pump mode is relevant, so larger SHG bandwidths are not helpful as long as the pump mode can be aligned to the SHG conversion band. Another aspect to consider in this comparison is that SHG is most applicable as a component of a system. Therefore, relevant SHG technologies should either support versatile and low-loss coupling techniques or be compatible with heterogeneous integration \cite{Chang2018,Stanton2019}. For example, to add SHG functionality to photonic integrated circuit (PIC) based on mature and scalable Si photonics, it is advantageous to either deposit, grow, or bond the nonlinear material to the Si substrate \cite{Stanton2019,Chang2019-2}.

Integrated $f$-$2f$ self-referencing without an optical amplifier \cite{Volet2017} or an assist-laser requires an SHG conversion efficiency on the order of 40~W$^{-1}$. At this level, 5~$\upmu$W of CW pump light (the typical power from a single comb line \cite{Li2017,Pfeiffer2017,Moille2019}) can be converted to 1~nW of signal light. To our knowledge, no single pass device has previously demonstrated this. While some resonant devices have exceeded this efficiency \cite{Bruch2018,Lu2019}, the narrow bandwidth and sensitive phase-matching condition makes them impractical for a fully integrated $f$-$2f$ self-referencing system. These resonant devices are better suited for parametric down-conversion applications with narrow bandwidths and squeezed light generation \cite{Luo2017,Guo2017,Bruch2018,Lu2019}. Furthermore, producing such systems on a large scale requires a high-yield and reproducible fabrication. This work describes an SHG device with the high efficiency and bandwidth suitable for an integrated $f$-$2f$ self-referenced comb. The high second-order nonlinearity of GaAs paired with the high index contrast and uniformity of integrating GaAs-based waveguides on Si has enabled this advancement \cite{Chiles2019}. We use birefringent modal phase-matching \cite{Chang2018,May2019} in a GaAs waveguide to produce SHG from a 2.0~\micron{} pump with an efficiency of 40~W$^{-1}$ and a signal bandwidth of 0.5~nm. The device is fabricated at the 76~mm (3-inch) wafer scale and is compatible with integration on the SiN platform used for the $f$-$2f$ self-referencing \cite{Stanton2019}.

In this manuscript, we present an optimized design for efficient SHG conversion with modal phase-matching and we discuss processing techniques for wafer-scale fabrication. The conversion efficiency is characterized to determine the peak efficiency, bandwidth, and temperature dependence. Finally, we conclude with a summary of the results and present an outlook for improving the SHG conversion efficiency.

\section{Design}

To design the SHG waveguide we select a phase-matched geometry that maximizes the conversion efficiency \cite{Yariv1973,Pliska1998}, $\eta$, defined as:

\begin{equation}
\eta = \frac{P_{2\omega}(L)}{P^2_{\omega}(0)} = \frac{2 \omega^2 \xi \kappa L^2}{n_{\omega}^2 n_{2\omega} \epsilon_0 c^3 } \,,
\label{eq:1}
\end{equation}

\noindent{}where $P_{2\omega}$ is the signal power, $P_{\omega}$ is the pump power, $L$ is the waveguide length, $\omega$ is the angular frequency of the pump, $\epsilon_0$ is the permittivity of free space, and $c$ is the speed of light in vacuum. The effective refractive indices of the pump and signal optical modes are $n_{\omega}$ and $n_{2\omega}$, respectively. The $\xi$ term in Eq.\,\ref{eq:1} accounts for phase-matching and the propagation loss:\footnote{Note that in \cite{Pliska1998} a typographical error exists: in Eq.\,3, the $\alpha$ appearing in the cosine term should be replaced with $\Delta$.}

\begin{subequations}
\begin{align}
\xi &=  \frac{\mathcal{A}^2 + \mathcal{B}^2}{(\Delta \alpha^2 + \Delta \beta^2)(L/2)^2} e^{ -( \alpha_{2\omega}/2 + \alpha_{\omega} ) L } \,,\\
\intertext{where}
\mathcal{A} &= \sinh{\left( \Delta \alpha L/2 \right)} \cos{\left( \Delta \beta L/2 \right)} \,,\\
\mathcal{B} &= \cosh{\left( \Delta \alpha L/2 \right)} \sin{\left( \Delta \beta L/2 \right)} \,.
\end{align}
\label{eq:2}
\end{subequations}

\noindent{}Parameters $\alpha_{\omega}$ and $\alpha_{2\omega}$ are the propagation loss coefficients for the pump and signal. The phase-mismatch, $\Delta\beta$, and the loss-mismatch, $\Delta\alpha$, are defined as $\Delta\beta = \beta_{2\omega} - 2\beta_{\omega} = (2\omega/c)( n_{2\omega} - n_{\omega} )$ and $\Delta\alpha = \alpha_{2\omega}/2 - \alpha_{\omega}$. A nonlinear coupling parameter, $\kappa$, in Eq.\,\ref{eq:1} accounts for two effects: the mode overlap between the pump and signal light and the interaction between the optical modes and the nonlinear optical susceptibility of the waveguide. It is defined as:

\begin{equation}
\kappa = \frac{\left| \iint_{\mathbb{R}^2} d_x E_{x, \omega}^2 E_{y, 2\omega} \,\mrm{d}x \,\mrm{d}y \right|^2}{\left( \iint_{\mathbb{R}^2} | E_{x, \omega} |^2 \,\mrm{d}x \,\mrm{d}y \right)^2  \iint_{\mathbb{R}^2} | E_{y, 2\omega} |^2 \,\mrm{d}x \,\mrm{d}y } \,,
\label{eq:3}
\end{equation}

\noindent{}where $d_x$ is the spatially varying effective second-order nonlinear coefficient corresponding to components of the $d$ tensor that interact with $x$-polarized pump light. The $x$-polarized component of the pump mode is $E_{x, \omega}$, and the $y$-polarized component of the signal mode is $E_{y, 2\omega}$. Wave propagation is in the $z$-direction and the waveguide top-surface in the $y$-direction. Each integral in Eq.\,\ref{eq:3} is taken over all space in the $x$-$y$ plane.

\begin{figure}[htbp]
\centering
\includegraphics{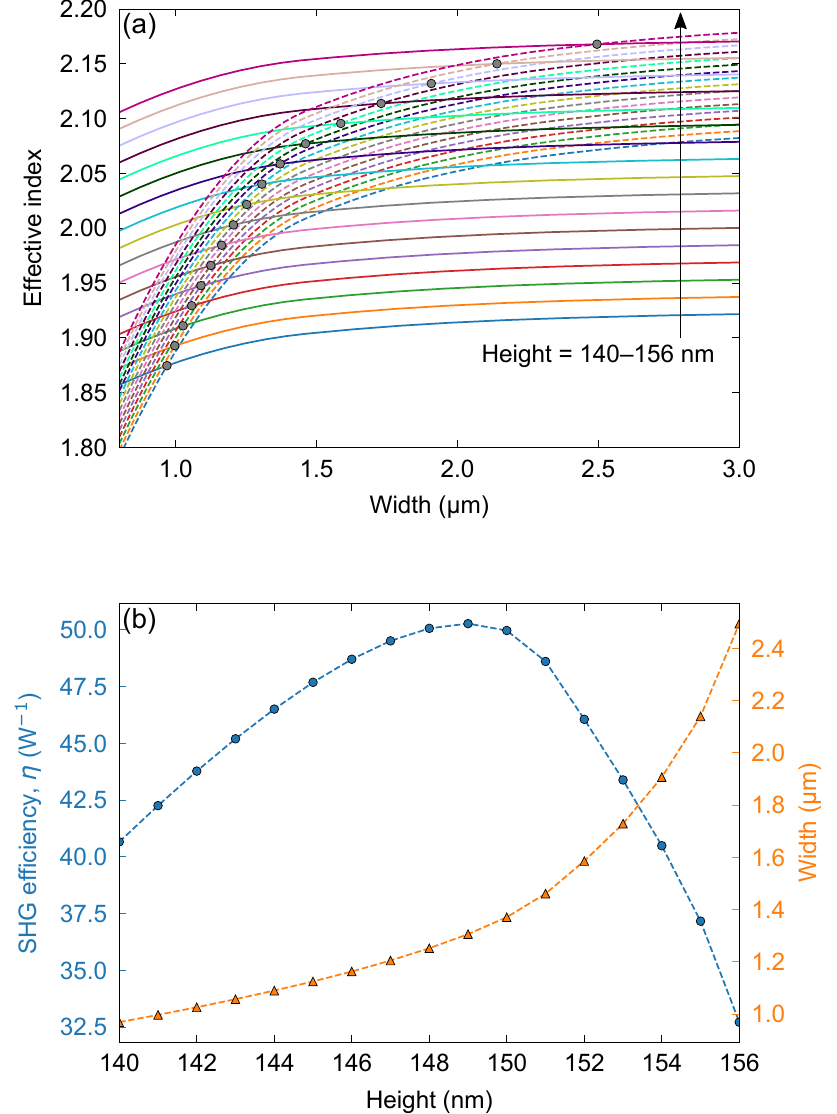}
\caption{(a) Effective indices of the pump (solid lines) and signal (dashed lines) as a function of waveguide width. Each curve represents a different waveguide height, from 140~nm with the lowest indices to 156~nm with the highest indices, in steps of 2~nm between curves. (b) SHG conversion efficiency ($\eta$) for a length of 2.9~mm plotted in circles on the left axis as a function of waveguide height for the perfectly phase-matched widths plotted in triangles on the right axis.}
\label{fig:PM}
\end{figure}

The magnitude of the $d_{14}$ and $d_{36}$ nonlinear coefficients of GaAs are equivalent for a $( 1 0 0 )$ oriented top surface with wave propagation in the plane of the $( 0 \bar{1} \bar{1} )$ primary flat or the $( 0 \bar{1} 1 )$ secondary flat. However, there is a wide range of reported values for this coefficient, and these values have large uncertainties because of the measurement techniques and differences in material preparations \cite{Choy1976,Eckardt1990,Roberts1992,Shoji1997,Skauli2002}. For the calculation in this work, we use $d_x = 180$~pm/V for GaAs, and we neglect its value in SiO$_2$ and air. Guided optical modes are calculated using a finite-difference method eigenmode solver from which $\kappa$ is evaluated by Eq.\,\ref{eq:3}. The GaAs is partially etched so a 15~nm thick slab remains, and it is cladded with SiO$_2$ on bottom and air on top. The effective refractive indices, $n_{\omega}$ and $n_{2\omega}$, and the nonlinear coupling, $\kappa$, are calculated for different heights and widths of the waveguide core. A Sellmeier model is used for the index of each material \cite{Skauli2003,Malitson1965}. The phase-matched geometries are found from the intersection of the discretely calculated values for $n_{\omega}$ of the transverse-electric (TE) polarized pump and $n_{2\omega}$ of the transverse-magnetic (TM) polarized signal. Similarly, $\kappa$ is found by interpolation. Figure\,\ref{fig:PM}(a) shows the effective indices of the pump and signal light for a range of heights and widths, and Fig.\,\ref{fig:PM}(b) shows the conversion efficiency $\eta$ for a length of 2.9~mm (the length used in this experiment) with the corresponding phase-matched geometries. The peak simulated conversion efficiency is $50.3$~W$^{-1}$, which occurs at a GaAs thickness of 149~nm and a width of 1.3~\micron{}. The pump and signal propagation losses used in this simulation are 1.5~dB/cm and 16.8~dB/cm, respectively.

\begin{figure}[htbp]
\centering\includegraphics{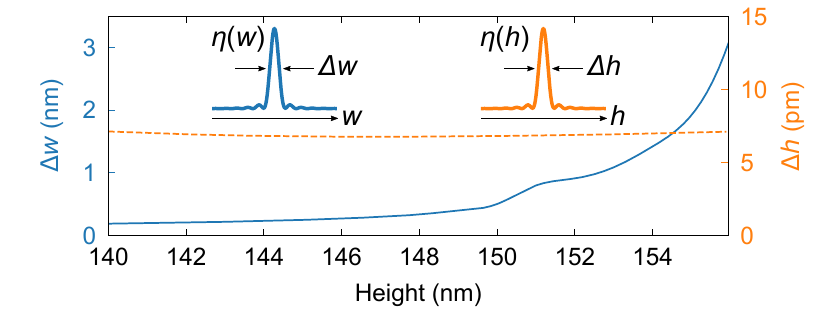}
\caption{FWHM of the conversion efficiency ($\eta$) for variations in the width ($\Delta w$) shown in solid blue on the left axis and the height ($\Delta h$) shown in dashed orange on the right axis.}
\label{fig:FWHM}
\end{figure}

Phase error between the pump and signal light due to width and height variations of the waveguide degrade the signal and typically limit the conversion efficiency of waveguided SHG devices. This effect is analyzed using the effective indices in Fig.\,\ref{fig:PM}(a). The FWHM bandwidths of the SHG signal are calculated for variations with waveguide width ($\Delta w$) and height ($\Delta h$), which are plotted in Fig.\,\ref{fig:FWHM} for a length of 2.9~mm. For larger heights, $\Delta w$ increases, indicating greater tolerance to width variations. However, the trend for $\Delta h$ is relatively constant for this range of waveguide heights. Although the peak conversion efficiency may be lower, an SHG waveguide with a height greater than 149~nm is less sensitive to geometric variations. For longer waveguides, these phase errors accumulate and the conversion efficiency becomes more sensitive to geometric variations. Our design targets a thickness of 154~nm for the SHG experiment, which has a phase-matched width of 1.9~\micron{} and a maximum possible conversion efficiency $40.5$~W$^{-1}$, as shown in Fig.\,\ref{fig:PM}(b).

The input facet waveguide width is 490~nm to minimize the coupling loss of the pump mode from a lensed fiber, which is simulated to be $3.6$~dB. This waveguide tapers to the SHG waveguide width over a length of 100~\micron{}. At the output facet the waveguide width is the same as the SHG section. The wide waveguide at the output reduces the facet coupling uncertainty compared to a tapered waveguide and the input facet taper is sufficient to suppress unwanted internal reflections. The simulated efficiency for the output coupling of the signal mode is $7.7$~dB, however, even higher loss is expected because this calculation does not account for higher-order mode coupling in the output fiber.

While GaAs and AlGaAs are ideal materials for SHG because of their large second-order nonlinearity, two factors limit their performance and application space. First, the bandgap of GaAs limits the SHG signal wavelength to $>$900~nm. For AlGaAs, shorter wavelengths down to $\sim$600~nm can be achieved depending on the Al mole fraction. The second limiting factor is the surface defect-state absorption \cite{Fiore1998,May2019}. Specifically, As-As bonds formed at the surface create a strong absorption feature with a peak near 950~nm \cite{Kaminska1987,Guha2017}. For GaAs, this absorption feature limits the minimum signal wavelength to $\sim$950~nm. Because of the larger refractive index, GaAs supports higher confinement and smaller modes compared to AlGaAs compounds. Therefore, we use GaAs since its bandgap is compatible with our application to convert $\sim$2.0~\micron{} light to $\sim$1.0~\micron{}.

\section{Fabrication}

\begin{figure}[htbp]
\centering\includegraphics{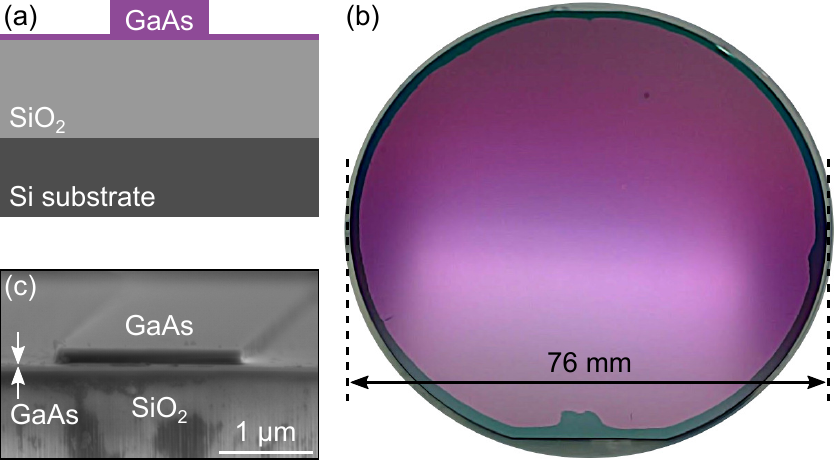}
\caption{(a) Fabrication process showing the bonding, GaAs substrate removal, and waveguide etch steps. (b) Picture of the 76~mm wafer shows nearly perfect yield of the transferred GaAs film. (c) SEM showing the output waveguide facet.}
\label{fig:fab}
\end{figure}

Figure~\ref{fig:fab}(a) shows a schematic of the designed waveguide cross-section. The fabrication process begins with molecular beam epitaxial growth on the (100) surface of a GaAs wafer. A 150~nm thick Al$_{0.8}$Ga$_{0.2}$As etch-stop layer is grown first, followed by the 158~nm thick GaAs waveguide layer. The waveguide layer is 4~nm thicker than is intended for the fabricated waveguide to account for material loss in subsequent HCl and NH$_4$OH steps. This GaAs wafer is then bonded to a Si wafer with a 3.0~\micron{} thick thermal oxide layer. Before bonding, both wafers are cleaned with solvents and activated with an atmospheric plasma containing metastable He and free-radicals of H and N. The bond is initiated in air at room temperature using a custom-built wafer bonding apparatus \cite{Bonder2019}, and an anneal at 150~\celsius{} on a hotplate for 1~hour completes the bond. Separate bonding experiments are used to measure the bond energy of $(0.91 \pm 0.16)$~J/m$^2$ with the double cantilever beam method \cite{Vallin2005}. While this is lower than other bonding recipes using SiN or Al$_2$O$_3$ interlayers, this direct bond minimizes the waveguide loss and is strong enough to complete the waveguide fabrication.

The GaAs substrate is removed in two steps. First, the bulk of the substrate is etched with a mixture of 350~mL H$_2$O$_2$ (9.8~mol/L) and 25~mL NH$_4$OH (14.5~mol/L) using an N$_2$ bubbler to stabilize the reaction. After the 150~nm thick Al$_{0.80}$Ga$_{0.20}$As etch-stop layer becomes visible (typically after 80~minutes), the wafer is transferred to a second mixture (also using an N$_2$ bubbler) with the same H$_2$O$_2$, but only 10~mL NH$_4$OH. Residue is removed with HCl (2.0~mol/L) for 20~s, and the etch-stop is selectively etched with HF (1.4~mol/L) for 30~s. Residual fluoride compounds produced from the HF etch are removed with \textit{N}-methyl-2-pyrrolidone at 80~\celsius{} for 5~min and then NH$_4$OH (1.3~mol/L) for 1~min. An image of the Si wafer at this point in the process with the transferred GaAs film is shown in Fig.~\ref{fig:fab}(b). The RMS surface roughness, measured with atomic force microscopy, on top of the GaAs is $\sim$0.2~nm over 9~\micron{}$^2$, which is the same as the bottom surface before bonding. The thickness uniformity is measured with optical interferometry to be $\pm$0.2~\% or $\pm$0.3~nm within a 20~mm radius from the wafer center.

The GaAs waveguides are defined with electron-beam lithography and inductively-coupled plasma (ICP) reactive-ion etching (RIE). The etch uses BCl$_3$ with 200~W ICP power, 100~W bias power, 5~mTorr chamber pressure, and a temperature of 20~\celsius{}. By dividing the etch in two steps and measuring the GaAs thickness in between, we consistently leave a 15~nm thick GaAs slab with an error of $\pm$1~nm, limited by the optical interferometric measurement. This GaAs slab is important to prevent exposure of the GaAs/SiO$_2$ interface to HCl, which etches the interface non-uniformly, causing phase-errors. The SiO$_2$ facets are etched with ICP-RIE using CHF$_3$ and O$_2$ gases, and the chips are released from the wafer using a deep-RIE etch of the Si substrate. A scanning electron micrograph (SEM) of the waveguide at the chip facet is shown in Fig.~\ref{fig:fab}(c). We estimate the sidewall roughness as 4.5~nm RMS using a top-view SEM. Finally, the chips are cleaned with HCl (1.1~mol/L) for 10~s to minimize the As-As bonds.

\section{Experiment}

\begin{figure}[htbp]
\centering\includegraphics{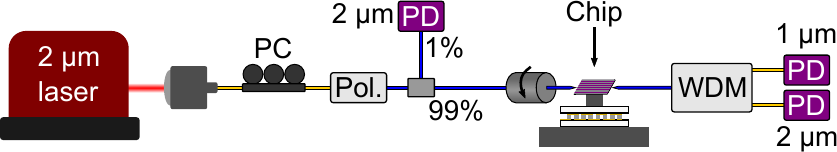}
\caption{Experimental setup for SHG. The yellow connections are single-mode fibers and the blue connections are polarization-maintaining single-mode fibers. PC: polarization controller; Pol.: fiber-based linear polarizer; PD: photodetector; WDM: wavelength division multiplexer for splitting the 1~\micron{} and 2~\micron{} light.}
\label{fig:setup}
\end{figure}

To measure the SHG conversion efficiency, we use a CW and tunable 2~\micron{} laser to pump the waveguides in the TE-polarized mode, as shown in Fig.\,\ref{fig:setup}. A $\sim$1~\% splitter is used to monitor the input power, and lensed fibers couple light on and off the chip. After the output fiber, a wavelength-division multiplexer (WDM) splits the TM-polarized signal and TE-polarized pump light so they can be monitored on separate photodetectors. Similar setups are used to measure transmission at 980~nm, 1064~nm, and 1220~nm using lasers at these discrete wavelengths.

\begin{figure}[htbp]
\centering\includegraphics{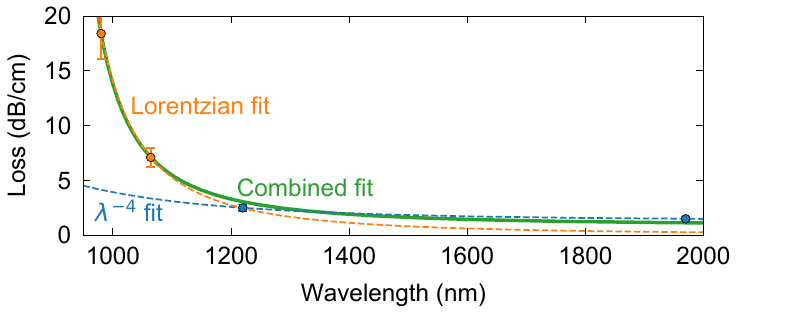}
\caption{Measured propagation loss showing the trends of a Lorentzian (dashed orange), $\lambda^{-4}$ (dashed blue), and a linear combination of both (solid green).}
\label{fig:broadband}
\end{figure}

Propagation losses measured near the signal and pump wavelengths are shown in Fig.\,\ref{fig:broadband}. We fit the loss spectrum to a linear combination of a Lorentzian function and $\lambda^{-4}$ to exemplify two separate spectral trends: the absorption feature at 950~nm and the scattering loss. The waveguide loss and facet coupling for both the pump and signal modes are characterized from transmission measurements of waveguides with different lengths on a single chip. Since bends are used to create the varying waveguide lengths, a second test-structure is used to verify that the bend loss are negligible. Pump light coupling loss at the input is $(5.0 \pm 0.5)$~dB and signal light coupling loss at the output is $(12.1 \pm 0.5)$~dB. The increased coupling loss compared to the simulation is likely caused by residue from the die-release process, observed in SEM imaging. Additionally, the signal light suffers from increased coupling loss due to the higher order modes in the output lensed fiber. The propagation loss for the pump mode is $(1.5 \pm 0.2)$~dB/cm and for the signal mode is $(16.8 \pm 1.8)$~dB/cm.

\begin{figure}[htbp]
\centering\includegraphics{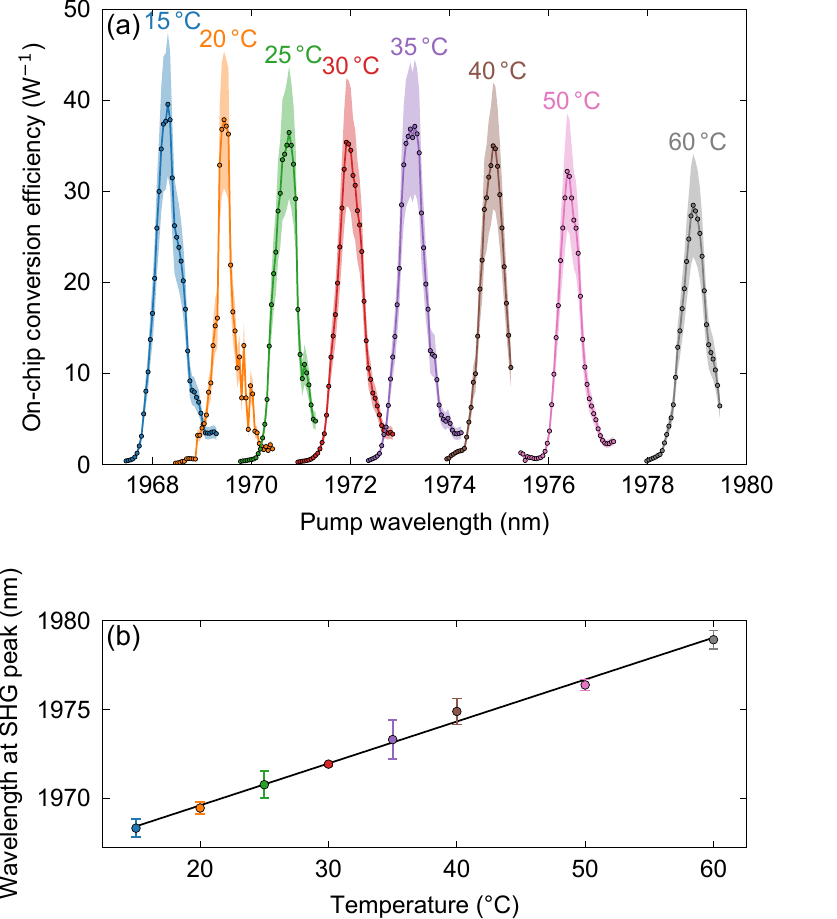}
\caption{(a) SHG spectra for various temperatures, limited by the temperature control setup. The standard error is indicated by the shaded areas. (b) Peak SHG wavelength for various temperatures with a linear fit of 0.236~nm/\celsius{}.}
\label{fig:shg}
\end{figure}

Conversion efficiency spectra are shown in Fig.\,\ref{fig:shg}(a) for a length of 2.9~mm and a range of temperatures. The peak efficiency is $(39.5 \pm 7.9)$~W$^{-1}$ at 15~\celsius{} and decreases to $(28.5 \pm 5.7)$~W$^{-1}$ at 60~\celsius{}. The uncertainty is calculated from the coupling loss variation across the chip. The change in peak SHG wavelength with temperature is 0.236~nm/\celsius{}, as shown in Fig.\,\ref{fig:shg}(b). The wavelength uncertainty arises from mode-hopping in the laser, and it is calculated from repeated measurements. We also measured SHG at 25~\celsius{} from devices with lengths of 6.3~mm and 9.7~mm, but the efficiency decreases to $(31.4 \pm 6.2)$~W$^{-1}$ and $(15.2 \pm 3.0)$~W$^{-1}$, respectively.

\section{Discussion}

Phase errors resulting from thickness non-uniformity likely limit the nonlinear conversion for the longer waveguides that produce lower conversion efficiency. By extrapolating the uniformity measurement from across the wafer to within the length of the SHG waveguide, we estimate that the average GaAs thickness varies by up to $\sim$20~pm. Compared to $\Delta h$ from Fig.\,\ref{fig:FWHM} of $\sim$7~pm, this thickness variation is very significant. We do not expect the average width of the waveguide to vary significantly compared to $\Delta w$ of $\sim$1~nm, though we have not been able to measure the width with this level of accuracy. The 2.9~mm long device may be more uniform than the maximum variation of $\sim$20~pm because it was located close to the center of the wafer, where the thickness variation is minimal.

Our key findings to reduce the waveguide loss at the signal wavelength involve minimizing As-As bonds at the GaAs surface at two different steps in the fabrication process. First, the bonding surface of the GaAs waveguide layer must be allowed to form a native oxide after the MBE growth, which is not removed before bonding. This direct bond between the native GaAs-oxide surface and the SiO$_2$ surface is optimal to reduce the As-As bonds. The second step is the surface treatment of the top and sidewall of the GaAs waveguide. Similarly, a native oxide is preferred, but only after an HCl strip of the surface reconstruction layer (the amorphous surface layer naturally formed in air) \cite{Parrain2015,Guha2017}. This allows for a native oxide to grow with minimal As-As bonds.

As indicated from the propagation loss spectrum, we have not completely avoided forming As-As bonds and the loss from this defect state can still be improved. A potential process to further reduce As-As bonds is to use atomic layer deposition (ALD) to passivate the surface \cite{Guha2017}. However, our attempts with ALD Al$_2$O$_3$ passivation produced higher loss than our current process. Also, the loss at the pump wavelength indicates that our sidewall roughness may have a significant contribution. Further optimization of our lithography and etching process may reduce the loss by decreasing scattering for both the signal and pump wavelengths.

At the elevated temperatures the decreased SHG conversion efficiency may be due to an increase in propagation losses. Free-carrier absorption in the GaAs for the signal is expected to increase with temperature \cite{Spitzer1959}, and the peak wavelength of the As-As defect state is strongly temperature dependent \cite{Kaminska1987}. The conversion efficiency does not degrade after the thermal cycling between 15~\celsius{} and 60~\celsius{}, so the loss is not a permanent effect. Regardless, we find that the conversion efficiency is consistent from 15~\celsius{} to 40~\celsius{}. An $f$-$2f$ self-referenced comb can reasonably be stabilized in this temperature range.

Similar SHG results were measured immediately following and several months after the fabrication, showing no measurable degradation in performance. The peak efficiency, bandwidth, and operation wavelength did not vary more than our measurement error within this time frame. These results show that the SHG performance is not affected by oxidation or other chemical reactions with the GaAs.

\section{Conclusions}

We present a 76~mm wafer-scale GaAs-on-insulator platform and an SHG device compatible with heterogeneous integration on a SiN PIC for $f$-$2f$ self-referencing and optical frequency synthesis \cite{Chang2018,Stanton2019}. Record-high SHG conversion efficiency of 40~W$^{-1}$ (476~W$^{-1}$cm$^{-2}$) in a 0.5~nm bandwidth is demonstrated from a single-pass device because of improvements made to the propagation loss and the waveguide uniformity. Efficient SHG operation is confirmed over a temperature range of 45~\celsius{}. With these new conversion efficiency achievements, it is now possible to realize a fully-integrated chip-scale $f$-$2f$ system. In future work to further improve SHG conversion efficiency, quasi-phase-matching would allow for optimization of the nonlinear mode-overlap and for implementing greater phase-matching tolerance. Applying orientation patterning to the GaAs platform is expected to produce even higher SHG conversion efficiencies, exceeding 100~W$^{-1}$ with a broader bandwidth of $>$10$^3$~GHz.

\section*{Funding}
This work was funded in part by the DARPA MTO DODOS program.

\section*{Acknowledgments}
We thank Kevin~L.~Silverman and Travis~M.~Autry for helpful discussions on the GaAs fabrication; Scott~B.~Papp for helpful discussions about applications; and Saeed~Khan, Abijith~S.~Kowligy, and Lynden~K.~Shalm for useful discussions and inputs on the manuscript. This is a contribution of NIST, an agency of the U.S. government, not subject to
copyright.

\section*{Disclosures}
The authors declare no conflicts of interest.


\bibliography{EJS_biblio}

\end{document}